# Increased genetic diversity improves crop yield stability under climate variability: a computational study on sunflower.

## Authors

Pierre Casadebaig (1)*, Ronan Trépos (2), Victor Picheny (2), Nicolas B. Langlade (3), Patrick Vincourt (3), Philippe Debaeke (1)
(1) INRA, UMR1248 AGIR, 31326 Castanet-Tolosan, France
(2) INRA, UR875 MIAT, 31326 Castanet-Tolosan, France
(3) INRA, UMR441 LIPM, 31326 Castanet-Tolosan, France
(*) Corresponding author

## Keywords

## Abstract

A crop can be represented as a biotechnical system in which components are either chosen (cultivar, management) or given (soil, climate) and whose combination generates highly variable stress patterns and yield responses. Here, we used modeling and simulation to predict the crop phenotypic plasticity resulting from the interaction of plant traits (G), climatic variability (E) and management actions (M).
We designed two *in silico* experiments that compared existing and virtual sunflower cultivars (*Helianthus annuus L.*) in a target population of cropping environments by simulating a range of indicators of crop performance. Optimization methods were then used to search for GEM combinations that matched desired crop specifications. Computational experiments showed that the fit of particular cultivars in specific environments is gradually increasing with the knowledge of pedo-climatic conditions. At the regional scale, tuning the choice of cultivar impacted crop performance the same magnitude as the effect of yearly genetic progress made by breeding. When considering virtual genetic material, designed by recombining plant traits, cultivar choice had a greater positive impact on crop performance and stability.
Results suggested that breeding for key traits conferring plant plasticity improved cultivar global adaptation capacity whereas increasing genetic diversity allowed to choose cultivars with distinctive traits that were more adapted to specific conditions. Consequently, breeding genetic material that is both plastic and diverse may improve yield stability of agricultural systems exposed to climatic variability. We argue that process-based modeling could help enhancing spatial management of cultivated genetic diversity and could be integrated in functional breeding approaches.

## Introduction

**Systems approach to crop improvement**

At the field scale, each component of the biotechnical system carries variability, whether it comes from plant genetic diversity, climate, soil or management actions.
Within a cropping landscape, canopies resulting from combinations of these elements in each field are subject to different intensities or timing of environmental stresses. Depending on these stress patterns, molecular and physiological processes (such as gene expression or plant growth) are differentially affected (Slafer, 2003 Tardieu, 2012, 2003 ). Consequently, farmers and plant breeders commonly observe that cultivars change in their relative performance across environments. Progress in crop improvement is limited by the ability to identify favorable combinations of genotypes (G) and management practices (M) given the resources available to experiment among possible combinations in the target population of environments (E) (Hammer and Jordan, 2007). Crop improvement can be viewed as a search strategy on a complex GEM space (Hammer and Jordan, 2007) to manage the genetic and environmental resources more efficiently by taking advantage of Genotype x Environment x Management (GEM) interactions.

**Exploring GEM combinations** *Phenotypic plasticity* is defined as the range of phenotypes a single genotype can express as a function of its environment (Nicotra et al., 2010). At the population level, the phenotypic plasticity resulting from the expression of genetic diversity in a *target population of environments* (Comstock, 1976; Cooper and DeLacy, 1994) has been referred as an *adaptation landscape* by Hammer et al. (2006). This landscape is only partly explored during the breeding steps, when phenotypes (mainly crop yield) measured in multi-environment trials and analyzed with various biometric methods (Cooper and DeLacy, 1994), are used to identify the overall most performing cultivars. Dynamic models based on crop physiology (process-based models) substitute the real system with an oriented and simplified view that can nevertheless be numerically explored to give a better insight on crop functioning by dissociating processes that closely interplay in the real system and that cannot be observed directly. Simulation has been proposed to complement and improve the resource-limited experimental exploration of the adaptation landscape (Chapman et al., 2003; Hammer et al., 2006; Messina et al., 2006, 2011) by partitioning environments at a continental scale (Chenu et al., 2013), searching for adapted traits for specific targeted environment types (Chapman et al., 2002) or evaluating cultivars with MET analysis methodology (Li et al., 2013).

For simulation aiming to be more relevant than using various indices built from crop or environmental data, dynamic models should present two properties.

Firstly, they should properly describe plant x environment interactions, by including feedbacks between physiological processes in the algorithm. For example, the water deficit computed from climatic data, as a function of precipitation and potential evapotranspiration (e.g aridity index) does not take account of the crop growth impact on evapotranspiration.

Secondly, descriptors of cultivar differences are needed among model inputs to predict phenotypic plasticity. These *genetic coefficients* (Jeuffroy et al., 2013; White and Hoogenboom, 2003) can take values either estimated by numerical optimization (Bannayan and Hoogenboom, 2009; Mavromatis et al., 2001) from model outputs, direct measurement of cultivar-specific phenotypic traits (Casadebaig et al., 2011; Villalobos et al., 1996) or genetic analyses (Quilot et al., 2005; Reymond et al., 2003).

**Searching for desirable phenotypes** Thinking the specifications of an ideal cultivar is similar to the *ideotype* concept which has long been used by breeders when screening plant populations. Before the advent of modelling approaches, some attempts to develop new ideotypes starting from the discovery of particular mutants and according to a priori approaches (ex: liguless or bm3 mutant for maize) were not so successful because the mutation showed negative counterparts.

Donald (1968) defined a crop ideotype as an ideal plant type with a specific combination of characteristics favorable for photosynthesis, growth, and grain production under a production environment based on knowledge of plant and crop physiology and morphology.

In the historical ideotype breeding approach, the term *ideotype* relates to phenotypes of individual plants. At the plant level, such an approach was implemented for crop improvement of barley (Rasmusson, 1987), maize (Mock and Pearce, 1975), rice (Peng et al., 2008) and wheat (Semenov and Stratonovitch, 2013; Sylvester-Bradley et al., 2012).

This concept has been later broadened to the canopy scale, by considering the effect of crop management (e.g on plant architecture Costes et al. (2013)). Various objectives were also specified, such as organic farming (Bueren et al., 2002) or pest management (Andrivon et al., 2013).

In this study, the design of crop ideotypes refers to the search of plant traits combinations that confer desirable properties at the crop level, given the climate variability and a chosen production situation (soil, management, genotype).

Simulation models are *de facto* embedded in the process. First, because the effect of climate variability on plant growth is tightly linked to the design of ideotypes. Second, because trait per trait analysis is very difficult to carry out through field experiments since even nearly-isogenic or mutant plants differ by multiple traits due to the feedbacks in plant functioning and/or pleiotropic functions of genes (Costes et al., 2013).

**Problem and aims**

We need to understand the magnitude, repeatability, and predictability of GE interactions to accommodate their effects (Vega and Chapman, 2006). For that, modeling and simulation are additional tools in the breeder's hand to measure and predict the phenotypic plasticity in a given population of environments.
In this study, we evaluate the possibility to design crop ideotypes with modeling and *in-silico* experiments. However, although this approach could be extended to other field crops, we focus on the sunflower crop as a case study. In France, sunflower is a rainfed crop growing in a wide range of soils regarding their potential water content available to the crop, while however mostly grown in areas submitted to a drought stress during grain filling period. In these conditions, agricultural advisory services wonder if they should consider locally adapted cultivars to potentially increase crop performance (Jouffret et al., 2011).
To tackle this question, we first built two virtual multi-environment trials (MET), based on either cultivated or *in silico* genetic diversity tested in a wide range of environments (including soils, climates and management practices). Then, we simulated indicators of crop performance (grain yield, oil concentration, radiation use efficiency) in these MET with a previously developed and validated model (Casadebaig et al., 2011). Ranking and optimization methods were used on these indicators to search for GEM combinations that match ideotype specification. Finally, we assessed performance gaps between designed ideotypes and an optimal use of current cultivars under different environmental context specifications (national, regional, local).

## Material and methods

**Model outline, parameterization and performance to simulate phenotypic plasticity**

**outline**  Computational experiments are based on the SUNFLO crop model, specifically developed to assess the performance of sunflower cultivars in agronomic conditions (Casadebaig et al., 2011; Lecoeur et al., 2011). This model is based on a conceptual framework initially proposed by Monteith (1977) and shared by numerous crop models as most of its terms are easily measurable on the field.
The crop growth rate is calculated with an ordinary differential equation [EQ1] where photosynthesis is described by an empirical conversion factor, the radiation use efficiency ($RUE$) (Monteith, 1994) ; $k$ being the light extinction coefficient and $PAR$ the photosynthetically active radiation.

[EQ1]
$$dDM/dt = RUE \cdot (1 - exp^{-k \cdot LAI}) \cdot PAR$$

Broad scale processes of this framework, the dynamics of leaf area index ($LAI$), light interception (radiation interception efficiency, $RIE$) and photosynthesis ($RUE$) were split into finer processes (e.g leaf expansion and senescence) to reveal genotypic specificity and to allow the emergence of interactions.

In cropping conditions, these processes are affected by numerous physical or biotic factors. Therefore, predictions with the SUNFLO model are restricted to attainable yield (Van Ittersum and Rabbinge, 1997) : only the main limiting abiotic factors (temperature, light, water and nitrogen) were considered in the model. These factors were modelled as multiplicative effects that limit the crop physiological processes (leaf expansion, transpiration, photosynthesis) described in potential growing conditions.

**input variables and parameterization**  The model inputs are split in four categories: genotype, soil and climate, management actions and initial conditions.

In the SUNFLO model, a cultivar is represented by a combination of 12 genetic coefficients whose values are assumed to be constant among environments, thus trying to mimic cultivar-specific genetic information (Boote et al., 2003; Jeuffroy et al., 2013). These genetic coefficients describe various aspects of crop structure or functioning: phenology (four phenostages), plant architecture (leaf area, number and vertical profile, canopy light extinction), response curve of physiological processes to drought (leaf expansion and stomatal conductance) and biomass allocation (potential harvest index and potential seed oil contration).

In our framework, a plant trait is considered as a genetic coefficient only if it (i) presents a significant phenotypic variability, (ii) has a stable value between phenotyping environment (similar ranking of genotypes), (iii) has a fair impact on the model output and (iv) is readily accessible by routine methods of measurement. The value of 10 of the 12 genetic coefficients are measured directly on field trials (Casadebaig, 2008; Debaeke et al., 2010) or specific experiments (Casadebaig et al., 2008). Even if these measured phenotypic traits are under an largely unknown genetic control, it was shown that the resulting phenotypic plasticity could be partially simulated (Casadebaig et al., 2011; Lecoeur et al., 2011).

In this study, we represented the genetic diversity by combining these genetic coefficients. These combinations could be observed as is, on commercial cultivars or could result from artificial arrangements (*virtual cultivars*).

Climatic data in field experiments were recorded by neighboring stations ($< 5$km) of the French meteorological network (Meteo-France) but precipitations were recorded at the plot level. Historic climatic data used for computational experiments came from the same source. Four climatic variables were recorded and computed at daily timestep : mean air temperature (°C, 2m height), global incident radiation (MJ/m2), potential evapotranspiration (mm, Penman–Monteith) and precipitation (mm).

Soil characteristics (texture and depth) and initial soil conditions (residual nitrogen, available water content) were obtained through soil core sampling and laboratory analysis. Management action (sowing date, planting density, irrigation, nitrogen fertilization) were collected from field experimenters

**performance** The model was recently evaluated on both specific research trials (Casadebaig et al., 2011; Lecoeur et al., 2011) and on trials that were more representative of its targeted use case, i.e. network of field trials for cultivar assessment (small scale MET, 16 sites x 20 cultivars) (Casadebaig et al., 2011). This evaluation showed that the model can rank cultivars in a MET. In research trials, climatic data recordings at the plot scale and soil granulometric analysis allowed to reduce uncertainty on model inputs (see supplementary figure 1 for a synthetic evaluation of the model). Depending on the uncertainty level in the inputs, the relative error for yield prediction was between 10 and 15%. However, this evaluation also revealed that the model could not discriminate between close-performing cultivars ($< 0.2$ t/ha).

A variance analysis of real and simulated networks indicated that genotype x environment interactions were significant both in the real and simulated networks, but with a much lower amplitude in the latter (mean square for GxE interactions was nine times weaker) (Casadebaig et al., 2011).

**Design of field experiments for genetic coefficients acquisition**

In this study, 53 sunflower commercial cultivars that were released between 1996 and 2012 were characterized to get genetic coefficients during eight field and six greenhouse experiments. Among the 12 genetic coefficients used in the crop model, eight were measured in the field experiments and two in greenhouse experiments (cf. figure 2). Greenhouse experimental protocols and related parameters measurements were previously described in Casadebaig et al. (2008). The two remaining coefficients (floral initiation and onset of senescence) were estimated as a linear function of flowering date.

| name | description | unit | reference |
| --- | --- | --- | --- |
| TDF1 | Temperature sum from emergence to the beginning of flowering | °Cd | (Lecoeur et al., 2011) |
| TDM3 | Temperature sum from emergence to seed physiological maturity | °Cd | (Lecoeur et al., 2011) |
| TLN | Number of leaves at flowering | leaf | (Lecoeur et al., 2011) |
| LLH | Rank of the largest leave of leaf profile at flowering | leaf rank | (Lecoeur et al., 2011) |
| LLS | Area of the largest leave of leaf profile at flowering | cm2 | (Lecoeur et al., 2011) |

| name | description | unit | reference |
|------|-------------|------|-----------|
| K  | Light extinction coefficient during vegetative growth | - | (Lecoeur et al., 2011) |
| LE | Threshold for leaf expansion response to water stress | - | (Casadebaig et al., 2008) |
| TR | Threshold for stomatal conductance response to water stress | - | (Casadebaig et al., 2008) |
| HI | Potential harvest index | - | (Casadebaig et al., 2011) |
| OC | Potential seed oil content | %, 0% humidité | (Casadebaig et al., 2011) |

Table 1: Description of the genetic coefficients used in the SUNFLO crop model.

Field phenotyping experiments were designed to generate optimal conditions i.e. to reveal the plant potential for development (high water and nitrogen input, deep soil) or for biomass allocation to grain (shallow soil to create moderate water stress at flowering then irrigation for optimal grain filling). These experiments were carried out one year after the cultivars were commercially released. Table 2 summarize the management and soil factors in these trials.

Field experiments were carried out on private land, under contract with CETIOM (Centre Technique Interprofessionnel des Oléagineux et du chanvre), no endangered or protected species were sampled. We used randomized complete block designs with 3 (Chambon location) or 4 (En Crambade location) repetitions of 30 m^2 (Chambon) or 39 m^2 (En Crambade) plots (6 ranks wide). Sowing at a doubled density was practised to target 6 plants per m^2 at emergence, after thinning. Trials were chemically protected to limit biotic factors that may affect crop functioning.

Crop phenology was regularly monitored, to identify the two key phenological stages (TDF1, TDM3) matching to the model parameters. By usual convention, a growth stage was defined as reached when 50% of the plants in the plot were displaying the adequate phenotype, using the phenological scale from CETIOM (2004). At flowering, the following variables were measured for five plants per block (15-20 plants per cultivar): total leaf number (TLN), area (LLS) and rank (LLH, from the stem base) of the largest leaf and plant height. Leaf area was estimated from measured values of leaf length and width, using a linear model (Casadebaig et al., 2008). At physiological maturity, 10 plants were sampled to measure harvest index (HI) as the ratio of dry grain weight to total dry aerial biomass. Grain yield was measured by harvesting of the four central rows and expressed on a dry basis (0% moisture) cleaned of trash. Grain oil content was determined by Nuclear Magnetic Resonance (Bruker NMS 110 Minispec NMR Analyzer) on 20 g dry achene samplings.

| location | year | cv # | sowing | harvest | dens. (pnt/ha) | res. N (kg/ha) | N fert. (kg/ha) | irrigation (mm) | capacity (mm) |
|----------|------|------|--------|---------|----------------|----------------|-----------------|-----------------|---------------|
| Chambon   | 2008 | 19 | 24/04 | 17/09 | 4.7 | 60.1 | 60 | 35  | 88.2 |
| Chambon   | 2009 | 20 | 21/04 | 09/09 | 6.5 | 60   | 60 | 105 | 68.6 |
| Chambon   | 2010 | 15 | 19/04 | 20/09 | 6.2 | 66   | 60 | 111 | 87.4 |
| Chambon   | 2011 | 19 | 20/04 | 15/09 | 6.5 | 72   | 0  | 74  | 87.4 |
| EnCrambade | 2008 | 19 | 09/04 | 01/09 | 6.3 | 170  | 53.6 | 0 | 252 |
| EnCrambade | 2009 | 21 | 12/05 | 03/09 | 6.3 | 92   | 80 | 0 | 252 |
| EnCrambade | 2010 | 15 | 19/04 | 01/09 | 5.6 | 214  | 14 | 0 | 252 |
| EnCrambade | 2011 | 19 | 02/05 | 01/09 | 5.6 | 135  | 57 | 0 | 252 |

Table 2: Summary table of the field experiments used to measure genetic coefficients.

**Design of computer experiments**

Two computer experiments were performed to compare two descriptions of genetic material growing in a similar target population of environments (cf. table 3).

For both experiments, *a priori* designs were built from factorial combinations of phenotypic traits values, pedo-climatic conditions and management decisions. Experiments differed only in the tested genetic diversity (G); the target population of environments was similar: 4560 Environment x Management (EM) conditions (cf. table 4). In the first experiment, the tested combinations of genetic coefficients were observed on a set of actual cultivars. Whereas, in the second one, virtual cultivars were created by recombining levels within the range of observed phenotypic traits.

| experiment | diversity | design method | studied factors | size |
|---|---|---|---|---|
| *real* | actual cultivars | factorial | genotype (8) | 36480 |
| *virtual* | virtual cultivars | factorial | genotype (2^5=32) | 145920 |

Table 3: Summary table of the computational experiments

**description of genetic diversity**  Among the 10 phenotypic traits used as genetic coefficients, a subset of five traits was chosen for the computational experiments. This choice was based on sensitivity analysis results (working with the most impacting traits) and the function of the trait in the model (avoiding direct multiplicative effects). Chosen traits (shaded distributions in figure 2) referred to plant phenology (the duration of grain filling phase), plant architecture (leaf area and profile) and environmental response (leaf expansion and conductance sensitivity to water stress).

In the first experiment (*real*), chosen combinations of trait values corresponded to existing cultivars. Among the 53 cultivars that were phenotyped for model parameterization, eight were used in this experiment. The choice for this subset was based on French cultivar sales (rank <= 3) in the last four years (2009-2012) (Mestries, pers. comm.). It is assumed that this experiment reflects broadly the actual genetic diversity of sunflower cultivated in France.

In the second (*virtual*), 32 virtual cultivars were created by combining 2 levels (min / max) for the 5 considered traits. The chosen levels were the extrema observed in the distributions of 53 cultivars (cf. figure 2). Besides this design on the 5 considered traits, genetic coefficients for the other traits were set as constant (mean value of the 8 real cultivars). The virtual cultivars were named with a 5 characters string of either lower or upper case, according to the level of the trait e.g. trait value for potential leaf area was coded *a* for small plants and *A* for large ones (cf. figure 3).

**target population of environments**  Five geographic locations (see figure 1) were chosen according to the main sunflower growing regions in France and different climate types (Joly et al., 2010; Metzger et al., 2005). Thirty-eight years (1975-2012) of weather recordings were used to represent the climate variability. This choice allowed to represent broad climate tendencies at the national scale : Mediterranean (*south* and *north* subtypes), oceanic (*Lusitanian*) and continental (*Atlantic central*) according to the typology of Metzger et al. (2005). The soil available water capacity was described according to three modalities, 100, 150, 250 mm to sample the diversity of soils present in each location.

The climatic variability between and within locations was assessed with indicators initially proposed by Whittaker (1970), i.e mean annual temperature and sum of annual rainfall (cf. figure 1). Locations were also characterized by calculating the aridity index on 38 years as the ratio between precipitation (P) to potential evapotranspiration (PET) (Middleton et al., 1997). The water supply does not satisfy the climatic demand in the locations where this index is below one. The aridity index in the target population of environments was mapped by using the CGIAR-CSI Global-Aridity and Global-PET Database (Zomer et al., 2008), where aridity index was computed on years 1950-2000.

Eight management options were also tested by combining two sowing dates (01 april; 30 april), two plant populations (5; 8 plants/m2) and two nitrogen fertilizer rates (0; 80 kg N/ha). The range of these management options is within observed practices in France.

Initial conditions were identical for all simulations in computational experiments: initial soil water content was set to full water capacity and residual nitrogen was 30 kg N/ha. The total size of the considered target population of environments is 4560 EM combinations (cf. table 4).

| component | factor | description | size |
|---|---|---|---|
| environment | location | growing regions | 5 |
| environment | year | historical climatic variability | 38 |
| environment | soil | soil available water capacity | 3 |
| management | sowing | sowing date | 2 |
| management | density | sowing density | 2 |
| management | nitrogen | nitrogen fertilization | 2 |

Table 4: Summary of the environmental and management factors considered to represent the target population of environments.

**Data analysis and simulation software**

**Data analysis strategy**  *Optimization analysis.* We first set up an optimization strategy to quantify difference in crop yield performance and stability between a global adaptation strategy (same cultivar all over the target population of environments) and a local adaptation strategy (optimal cultivar choice for each EM combination).

To perform this analysis, we defined a performance and a stability criterion, for nine given contexts of cropping conditions knowledge.

More formally, a context is a subset of the *{location, soil, management, year}* set. We focused on the first three variables (8 possible combinations), including the empty set (*global* context) that corresponds to the global adaptation strategy. One additional context was defined : *local* (i.e. *location:soil:management:year*) corresponding to the best possible local adaptation, including knowledge on the climatic conditions. Contexts and instances used in the study are detailled in table 6. The set of instances of a context, denoted $I(c)$, is formed by combining all possible values for each variable of the context. We defined the performance criterion for an instance $i$ by the maximal expected value of the yield obtained by selecting the most performing cultivar:

$$perf(i) = max(E[Yield|i])$$

To evaluate a context, we computed the mean value of this performance over all instances of a context.

$$perf(c) = \frac{\sum_{i \in I(c)} perf(i)}{\#I(c)}$$

Our aim was to estimate the maximal expected value of yield knowing that we consider for each instance of the context $c$ the cultivar that provides maximal yield.

Thus, the context that maximises the *perf* indicator should inform us on the subset of variables (amongst location, soil and management) that is the most relevant for choosing the cultivar.

Mathematically, the set of subsets of variables can be organized as a lattice (for lattice theory, see Grätzer (2003), first concepts and distributive lattices chapter) and choosing the best context consists in browsing this lattice using the *perf* indicator and the stability criterion described below.

The second criterion evaluated the stability of the strategy, relying on the standard deviation of the yield :

$$sd(c) = \sigma[Yield|cultivar = argmax(perf)]$$

Here $argmax(perf)$ stands for the cultivar that is selected for any instance of $c$ by maximising the $perf$ indicator.

For example, the context *c = {location, soil, ., .}* is considered to illustrate the optimisation algorithm. The 15 defined instances are *I(c) = {{soil=100 and location=Avignon}, {soil=100 and location=Toulouse}, ..., {soil=250 and location=Poitiers}}*. The algorithm first computes the mean values by cultivars for each combination of locations x soils (15 instances) i.e. we browsed location x soil combinations but not management x year combinations (fixed). The best cultivar is then selected for each instance, thus providing a list of optimized choices (15 cultivars, potentially different). The performance indicator for the context is the mean value of this vector. The stability indicator is finally computed by merging the previously identified optimal cultivars decisions with the simulated dataset ; i.e, for this *location:soil* context, each instance is assigned one single and fixed choice of cultivar for each of the remaining degrees of freedom (in this case, management x years combinations). This procedure sets a fixed number of observations across contexts (n=4650), allowing to compare stability indicators.

*Ranking analysis.* In a second step, we refined the previous analysis by running a Condorcet election method (Schulze, 2011) that creates a single sorted list of winner cultivars by comparing multiple ranking lists. Input data was generated by ranking cultivars for each location x soil x year x management (EM) combinations, on decreasing grain yield. This approach was performed only for two contexts that match current cultivar advisory practices: national (context *global*, one list) or regional (context *location:soil*, 15 lists).

**Simulation and data processing** The model development and simulations rely on the VLE software (Virtual Laboratory Environment) (Quesnel et al., 2009) and the RECORD platform (Bergez et al., 2013). This platform implements a mathematical specification (DEVS, Discrete EVent system Specification (Zeigler et al., 2000)) that was originally developed for research in modeling and simulation independently of the domain of application.

A range of indicators was computed to describe the system, based on simulation data (cf. table 5). Because the effects of abiotic stress on crop functioning were included in the model, it was possible to summarize the stress patterns experienced by each GEM combination. We defined indicators of water, temperature and nitrogen stress impact on the crop photosynthesis as the integration of the dynamic stress factors (output variables) over the cropping period (sowing - harvest).

| factor | indicator | unit | formula |
|---|---|---|---|
| water | Climatic water deficit | mm | sum(P - PET) |
| water | Edaphic water deficit (continuous) | days | sum(1 - FTSW) |
| water | Water stress impact on crop photosynthesis | days | sum(1 - f(FTSW)) |
| nitrogen | Nitrogen stress impact on photosynthesis | days | sum(1 - f(NNI)) |
| temperature | Thermal stress impact on photosynthesis | days | sum(1 - f(TM)) |
| - | Photosynthesis | g/MJ/m^2 | mean(RUE) |
| - | Grain yield | t/ha | - |
| - | Grain oil content | % | - |

Table 5: **Indicators of crop functioning computed from simulation data.** Output variables upon which indicators were computed : rain (P), potential evapotranspiration (PET), fraction of transpirable soil water (FTSW), nitrogen nutrition index (NNI), mean air temperature (TM), radiation use efficiency (RUE). Functions used as abiotic stress scalars are documented in (Casadebaig et al., 2011)

Experimental and simulated data were processed with R software (R Core Team, 2012) with R packages *plyr* (data processing) (Wickham, 2011), *agricolae* (biplots) (Mendiburu, 2013), *ggplot2* (visualization) (Wickham, 2009), *ez* (visualization of the correlation matrix) (Lawrence, 2013) and *knitr* (reporting) (Xie, 2013).

## Results

### Climatic and phenotypic data description

**Climatic variability**   Locations chosen for computational experiments undergo contrasted climates at the national scale but are within a same range of aridity index (cf. figure 1, panel A). Mean trends (over 1975-2012) for the five locations (cf. figure 1, panel B) indicated that the sum of annual rainfalls were between 800mm (oceanic climate) and 600mm (Mediterranean and continental). Mean annual temperature ranged between 10.5 °C (continental) and 14.4 °C (Mediterranean).

The narrow geographical sampling was compensated when looking at inter-annual variability (cf. figure 1, panel B). In the 190 locations x year population, 9% could be considered as semi-arid conditions and 13% as dry subhumid, according to Middleton et al. (1997) boundaries for dryland classification.

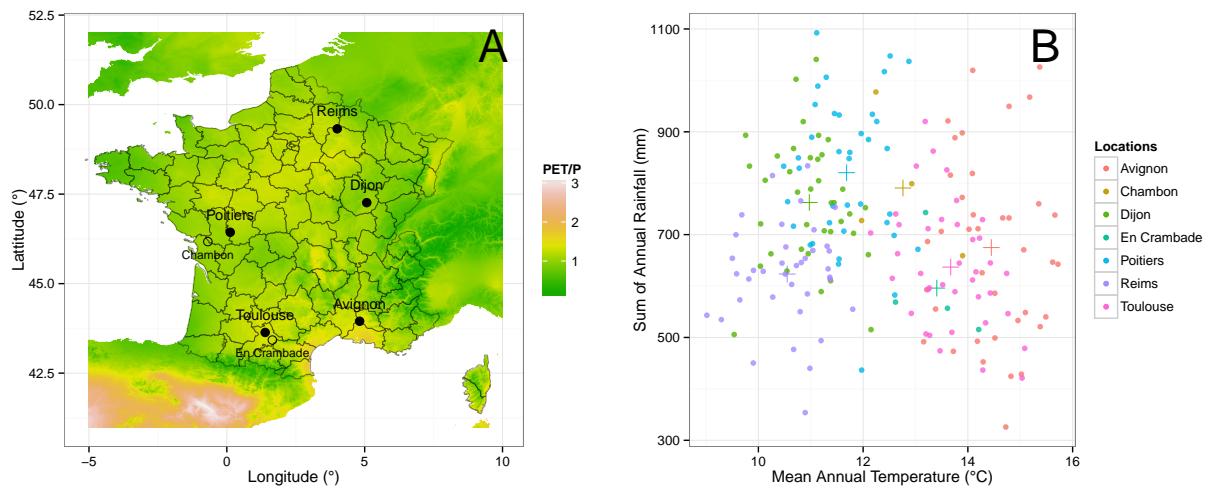

Figure 1: **Geographical locations and climatic variability in the target population of environments.** The left panel (A) displays a map of potential evapotranspiration ratio (1 / arididy index, data from (Zomer et al., 2008)), filled symbols represent locations used for computational experiments, open symbols represent phenotyping locations. The right panel (B) shows climatic variability with individual years (points) or mean (crosses) in locations used in simulation (historical variability over 1975-2012) and for parameterization (2008-2011).

**Genotypic variability**   Figure 2 displays the genotypic variability in the genetic coefficients, on 53 cultivars that were released between 1996 and 2011, with 50% of cultivars released after 2007. Correlations between the genetic coefficients were observed in the phenotyped material (figure 2). The most significant correlations were observed for coefficients related to architectural traits (*leaf number* and *leaf profile*). Correlations between response traits and other types of traits (architecture, development) were also significant, but less strong and are not physiologically relevant.

A preliminary study on a small factorial cross between inbred lines indicated that seven of the 12 traits studied have quite high heritabilities (Triboi et al., 2004)

### Prediction of the phenotypic plasticity in the target population of environments

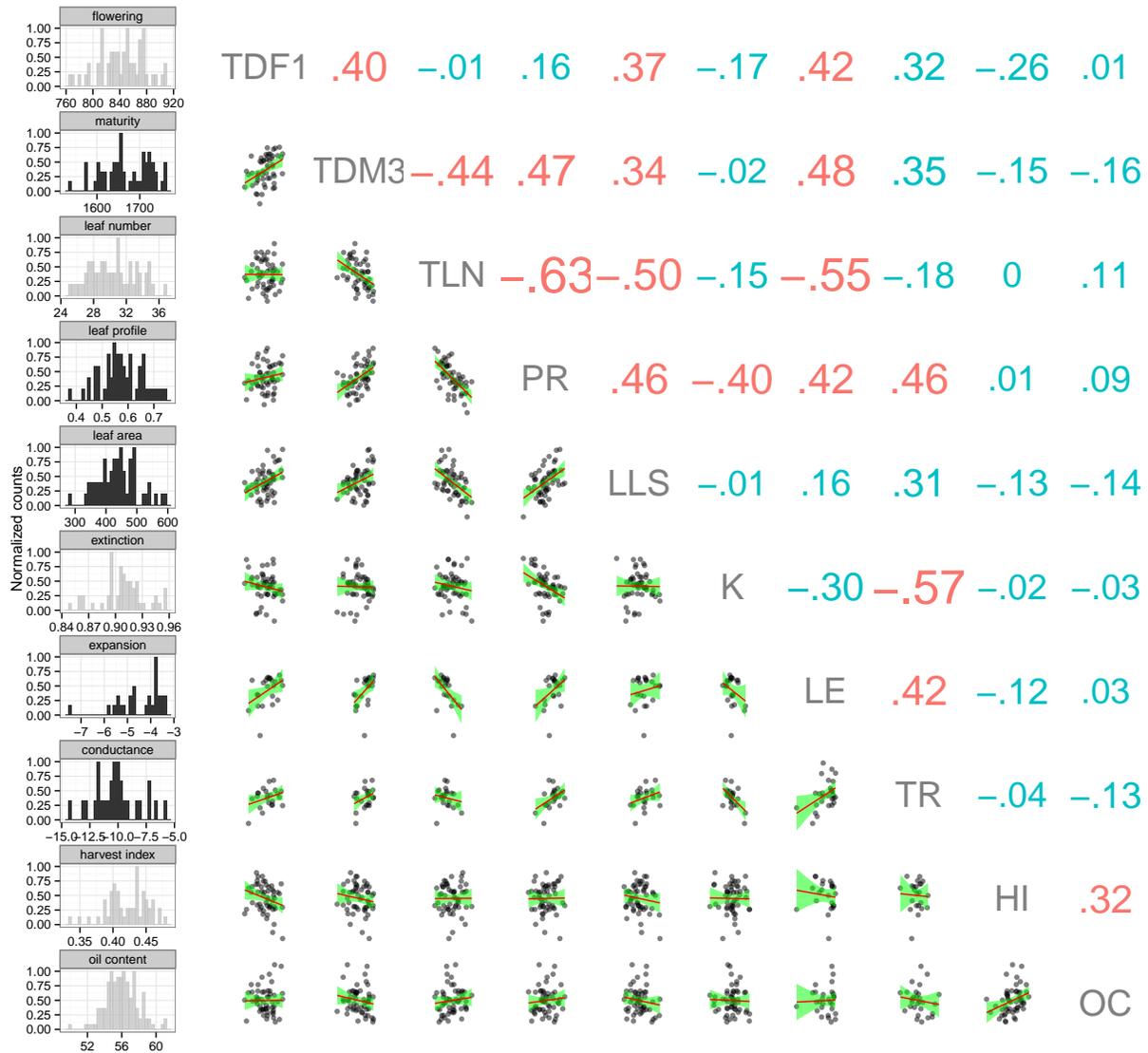

Figure 2: **Genotypic variability and correlation in the genetic coefficients.** Distribution of the values of genetic coefficients for 53 cultivars (left column). Visualization of the correlation matrix (right part): Pearson's correlation coefficients are displayed in the upper-half part, significant coefficients (Pearson's product moment correlation) are indicated in red.

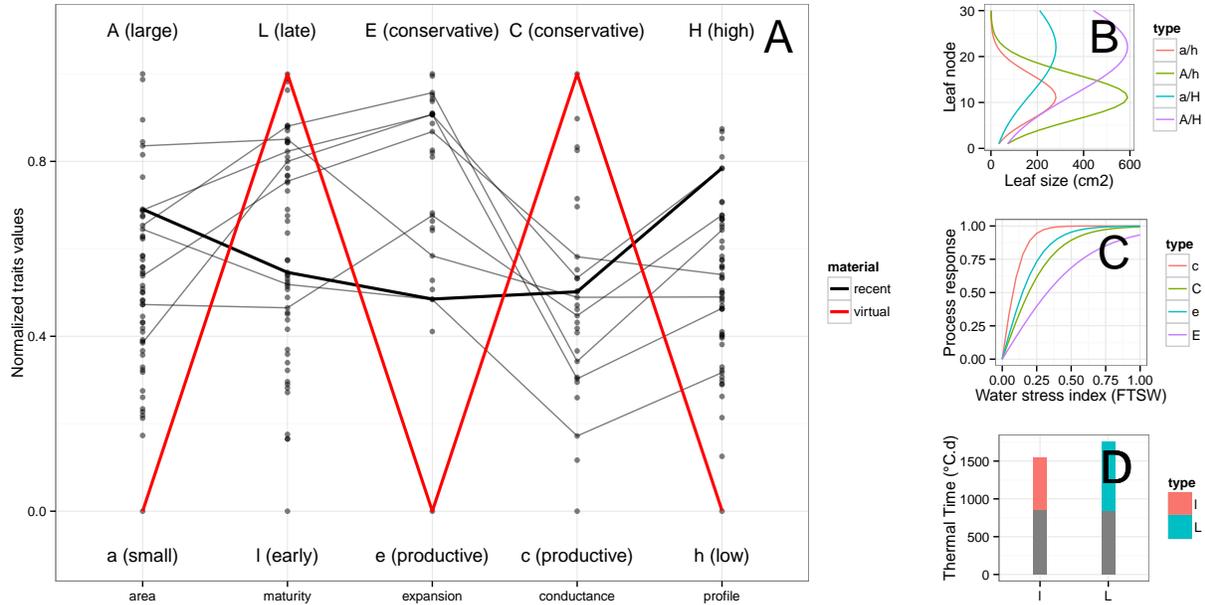

Figure 3: **Variation in phenotypic traits and their impact in crop model.** The left panel (A) displays the genotypic variability observed for five phenotypic traits in 53 commercial sunflower cultivars. Traits values were normalized to [0,1] according to their variation range and letters illustrate the coding used to name virtual cultivars. Line highlights specific trait combinations corresponding to recent (black, with cv. *Extrasol* in bold) or virtual genetic material (red, *aLeCH* combination) that were evaluated in computer experiments. The remaining panels show the function of these traits as described in the dynamic crop model, legends refer to the coding in panel A. Panel B focus on *area* and *profile* traits impact on crop architecture: the former modulates the plant potential leaf area and the later modulates its vertical distribution; all four possible combinations are displayed. Panel C focus on cultivar-specific leaf *expansion* and stomatal *conductance* sensitivity to water stress: the value of the response parameter modulates the shape of the response curve of the considered process. Panel D illustrates the impact of *maturity* parameter: its value modulates the duration of the grain-filling phase (flowering date being fixed).

**Visualization of simulated phenotypic plasticity**   Each individual GEM combination experienced a unique pattern of abiotic stress, summarized with indicators computed from simulation data and presented in table 5. As climatic variability was described by using temperature and water related indicators (cf. figure 1), we used the same factors to assess the variability of abiotic stress experienced by the GEM population.

Figure 4 displays the density of GEM population in a 2D space defined by water and thermal stress impacts on crop photosynthesis. This result allowed to visualize the phenotypic plasticity of different genetic materials growing under various climates. In other words, this representation was helpful to figure out the effect of environmental variability (climate and management) on a given population of cultivars by switching from a climate-based (in figure 1) to a crop functioning point of view.

For example, the computed water stress intensity (y-axis) was affected by rainfall on the crop cycle, soil water capacity and by the adaptative responses of the cultivar (leaf area growth, stomatal conductance). Mean water stress intensity was stronger and its variance in the GEM population was much narrow in drought-prone location (Mediterranean south climate type).

When considering the experiment with real cultivars (upper graphs), the presence of two high-density areas in some locations was the consequence of discrete levels of soil water capacity. These areas were not visible in virtual cultivars experiment, as the sampling method of genetic coefficients generates more diverse cultivars, leading to a more continuous response of the population.

The temperature stress indicator (x-axis) increases if mean temperature deviates for an optimal range for photosynthesis. Consequently, mean thermal stress intensity shifted among locations in relation to the crop thermal requirements : southern locations showed a better suitability by minimizing thermal stress. In this case, temperature was not high enough to affect crop performance. For this indicator, the high density areas were linked to sowing dates rather than cultivar precocity.

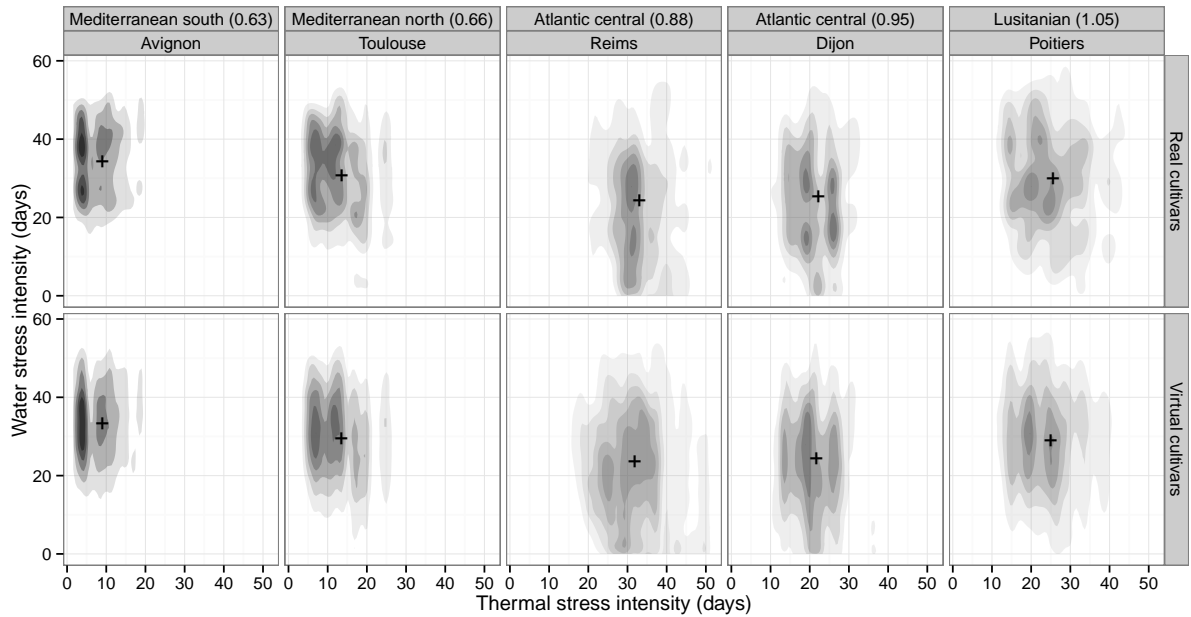

Figure 4: **Plasticity of GEM population in a space defined by abiotic stress, for two genetic materials.** Locations are sorted from left to right according to their aridity index (in brackets). The top row corresponds to real cultivar and the bottom row to virtual cultivars. In each panel, the density of GEM population is interpolated in a 2D space defined by indicators of cumulated water (y-axis) and thermal (x-axis) stresses impact on crop photosynthesis during the cropping duration (cf. table 5). Symbols (+) indicate the mean value for the population.

**Searching for crop ideotypes that optimize yield level and stability**  *Optimization approach.* By design, real and virtual genetic material had similar mean values for the phenotypic parameters which were not considered in design of experiments (grayed out distributions in figure 2). When comparing the global mean performance between genetic materials on the simulated dataset (cf. table 6), real and virtual cultivars had a similar level of performance (+0.014 t/ha, 0.55 % increase for virtual cultivars). When the same comparison was achieved on a dataset optimized for cultivar choice, virtual cultivars performed better than real ones (+0.14 t/ha, 5.3 % increase) ; even if this choice was made on a global basis (*global* context). When considering the cultivar choice level i.e. the difference between baseline and optimised scenario (the first two lines of table 6) indicated a more solid increase in performance with the virtual cultivars (+7.1 %) than with the real ones (+2.2 %).

| dataset | context | decision (#) | real (t/ha) | virtual (t/ha) | difference (t/ha) | ratio (%) |
|---|---|---|---|---|---|---|
| baseline | - | | 2.53 | 2.55 | 0.01 | 0.55 |
| optimized | global | 1 | 2.59 | 2.73 | 0.14 | 5.28 |
| optimized | soil | 3 | 2.61 | 2.73 | 0.12 | 4.74 |
| optimized | location | 5 | 2.59 | 2.73 | 0.14 | 5.47 |
| optimized | management | 8 | 2.59 | 2.73 | 0.14 | 5.46 |
| optimized | location:soil | 15 | 2.61 | 2.73 | 0.13 | 4.85 |
| optimized | location:management | 40 | 2.59 | 2.73 | 0.14 | 5.52 |
| optimized | soil:management | 24 | 2.61 | 2.73 | 0.13 | 4.9 |
| optimized | location:soil:management | 120 | 2.61 | 2.74 | 0.13 | 5.01 |
| optimized | local | 4560 | 2.62 | 2.79 | 0.17 | 6.3 |

Table 6: **Performance of real and virtual genetic material in relation with optimization of the cultivar choice.** Impact of cultivar adaptation strategies depending on cropping conditions knowledge (context) and genetic material (real and virtual). In column *dataset*, "baseline" indicates that crop performance was computed without optimizing the cultivar choice ; "optimized" refers to the optimization strategy implemented for different contexts. Column *decision* indicates the number of distinct cultivar choices (instances) that are evaluated for each context. Columns *real* and *virtual* are crop performance indicators (cf. data analysis section). Comparisons between genetic material were computed in *difference* (virtual - real) and *ratio* (virtual - real)/real columns.

When focusing on each genetic material, optimization globally granted a better crop performance with the detailed knowledge of environmental and management contexts (cf. figure 5). Difference in crop performance between global and local advisory strategy (*global* and *local* contexts) was 0.033 t/ha for real and 0.061 t/ha for virtual cultivars. However, performance gaps were not evenly distributed along context levels. For both genetic materials, but specially with virtual cultivars, performance was clearly increased when considering the most detailed context (*local*).

For real cultivars, the knowledge of soil water capacity (all contexts with *soil*) provided a major increase in performance (+55% of total impact). The gain in yield stability (as indicated by the 10th percentile) was weak with this genetic material. For virtual cultivars, performance gaps were more progressive along contexts : they regularly increased when adding elements in contexts. Unlike real material, yield stability increased more steadily with performance : gain in the 10th percentile was 0.05% with real material and 3.6% for virtual material. Without considering the *local* context, yield stability was increased when considering the soil potential water capacity in cultivar choice.

This difference in the stability response between genetic materials was explained by crop yield extremum. The optimization of cultivar choice has mainly improved the highest tier of performances (90th percentile) of real cultivars, whereas for virtual cultivars an increase is observed for both the lowest and highest tier (figure 5 illustrates this with the 10th percentile).

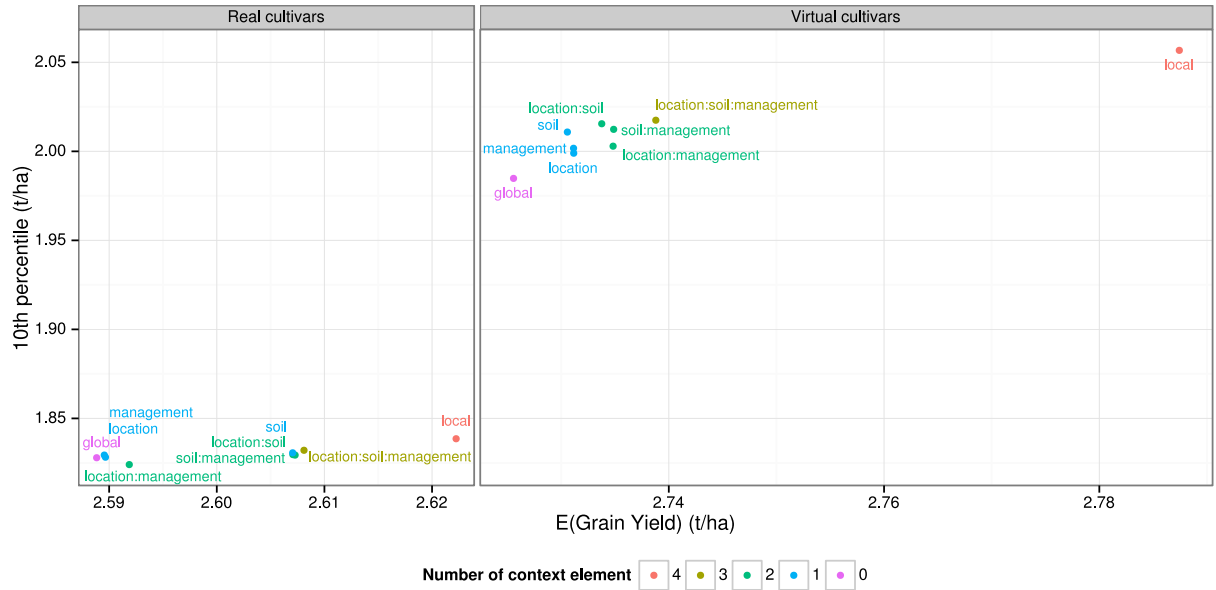

Figure 5: **Impact of cultivar choice on crop yield performance and stability for two genetic materials.** Points represent different levels of context in regard to the decision process ; color indicates the number of elements that defines a context. For the context *global*, cultivar choice is optimized for all EM combinations at once; for the context *local* this choice is optimized for each combination (table 5 indicates the number of choices being made).

*Ranking analysis.* The overall rankings produced by the election method (legends in figure 6) were different from ordering cultivars on mean grain yield value (as in the optimization analysis), as the algorithm choose the "least worst" cultivars. The overall winner for virtual cultivars (*aLeCh*) was different from the one with best mean yield (*ALECh*) and the two lists were poorly correlated (Kendall's tau = 0.06, p = 0.59). However, for real cultivars, election and optimization methods gave very similar results, with only an inversion at the 7th rank (data not shown).

For real cultivars, cv *Extrasol* displayed the best global adaptation ability (*global* context). This cultivar, released in 2007, obtained major market shares in France in 2010-2012 (Mestries, pers. comm.). The ranking at national level was not correlated (Kendall's tau = 0.14, p = 0.71) to the measured potential harvest index parameter of the cultivars (cf. table 1). As the simulated yield is strongly sensitive to this parameter, the lack of correlation indicated that rankings resulted from interacting parameters rather than an additive effect of this single plant trait closely related to yield.

At a regional level, it performed best for 9 out of 15 locations x soils tested, reflecting its adaptation potential. Its stability at the first rank was higher in Mediterranean (Avignon and Toulouse locations) and shallow soils (100 and 150 mm) than in deep soils, where it was outperformed by a better adapted cultivar (cv *Vellox*). Other elected cultivars were better adapted either for deep (cv *Vellox*) or shallow soils (cv *Biba*) but they also performed well at a national context. Remaining cultivars (below the 5th rank) did not performed well on any regional context.

For virtual cultivars, *aLeCh* combination of traits was the best overall ideotype, even if it performed best only in 6 out of 15 regional environments. Its frequency at the first rank (∼ 15 %) was lower than that of the second best overall virtual cultivar (*ALeCh*, ~ 40 %), meaning that it avoided low rankings but rarely got top ones. This ideotype had moderate leaf area development, late maturity, weak leaf expansion and strong stomatal conductance regulation during water deficit (see Casadebaig et al. (2008) for these response traits) and a low leaf area profile. All these traits fit in drought adaptation property : a plant with a relatively small leaf area, that regulates its water loss by lowering its conductance rather than reducing its leaf expansion. Additionally, as the largest leaves are down in the canopy, these are the first that are senescent, further lowering leaf area during terminal stresses. Its moderate photosynthetic capacity (18th place when ranked for mean RUE, cf. table 5) was compensated by a longer cycle duration.

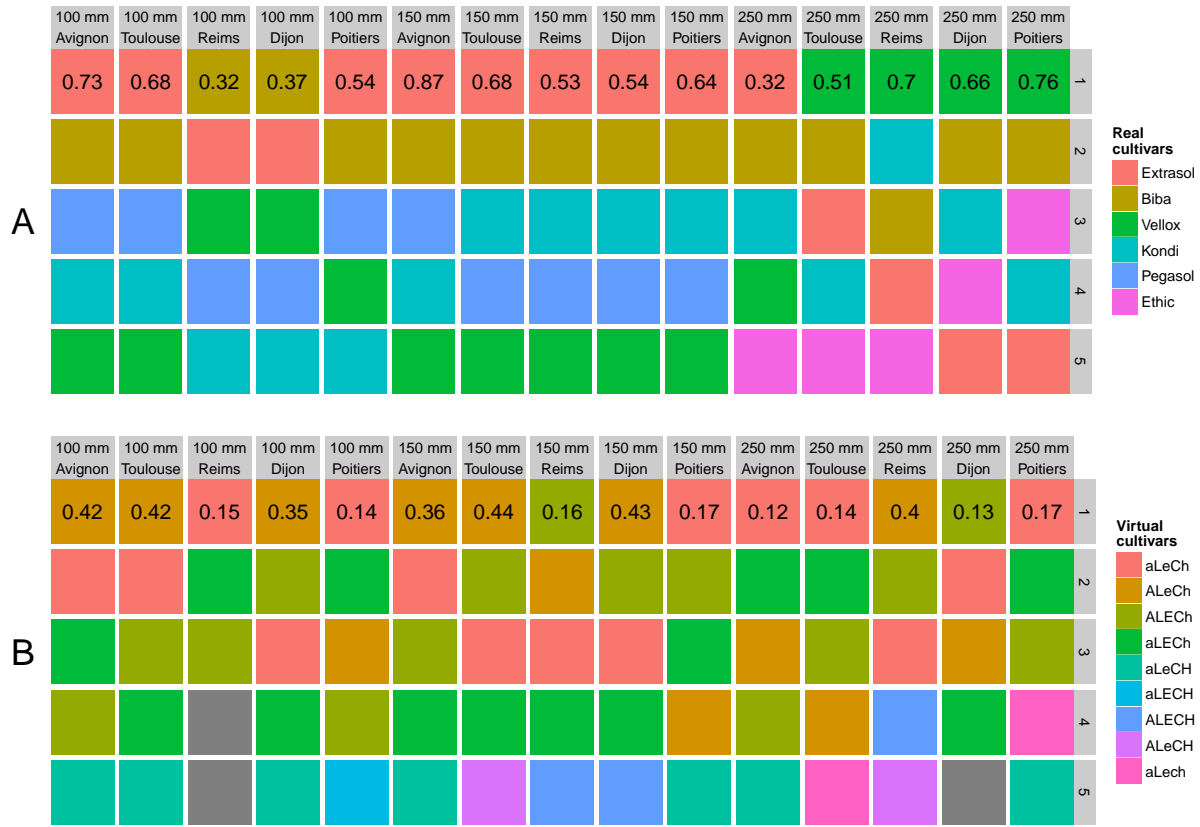

Figure 6: **Sorted lists of cultivars according to a Condorcet election method (Schulze, 2011) on simulated grain yield.** The upper panel (A) targets real cultivars, lower panel (B) is for virtual cultivars. In each panel, cultivar ranking (vertically from best to worse, right labels) is displayed in relation with regional environments (soil water capacity x location, horizontally) as in conventional GxE interaction plots. Overall ranking for cultivars is presented in legends. Gray cells are caused by ties in election method. The frequency at the first rank is displayed as an indicator of ranking stability within a regional context and over the remaining years x management cases.

The three other ideotypes displayed a similar set of traits, with variations either in potential leaf area (*ALeCh*) and/or leaf expansion (*ALECh, aLECh*). The conservative strategy of stomatal conductance regulation, late maturity and to a lower extent low leaf area profile were the most frequent traits among best ideotypes. These traits caused plastic response in plants : e.g. the variance for leaf area index was significantly higher ($p < 0.001$) with a more conservative water management strategy, meaning that this trait value was linked to a higher phenotypic plasticity.

The directions for desirable trait value were not similar when comparing the real and the virtual overall best cultivars (cf. figure 3, panel A). Two traits were specifically concerned : area (smaller plants) and leaf profile (lower leaves). However, when comparing the position of the overall best real cultivar among recent genetic material, its values for the reponse traits (*expansion* and *conductance*) were closer to boudaries pointed by optimization, relatively to recent material; i.e. lower values for the regulation of leaf expansion (late regulation under increasing water stress) and higher values for the regulation of stomatal conductance (early regulation).

## Discussion

**Application for plant breeding : toward agrogenomics**

Improving crop performance via plant breeding is a successful lever (on sunflower, +1.3% grain yield / year, see Vear et al. (2003)) and is mainly based on measuring phenotypic performance (mainly yield) of testcrosses of progenies from segregating genetic material growing in multi-environment trials.

In ideotype breeding, the relation between crop yield and its determining factors is implicitly modelled. In this case, an ideotype is basically a virtual, intermediary object (Vinck et al., 1995), useful to formulate, confront, and sometimes reconcile, the visions of specialists (Andrivon et al., 2013). In this case, a crop ideotype is mainly defined at the plant level, focusing on phenotypic factors.

Model-assisted breeding basically achieves similar objectives with tools needing a greater formalization (dynamic modeling, design method). Both of these tools are exchangeable either to test other models (Rosenzweig et al., 2013) or designing methods from scientific fields ranging from genetics (Cooper and DeLacy, 1994), ecology (Nicotra et al., 2010; Sadras et al., 2009) or computer science (Letort et al., 2008).

Our approach focus on using a process-based crop model and mathematical optimization methods to design an ideotype based on a set of phenotypic traits values, although model-assisted breeding can take different paths. This computational study compared real and virtual cultivars designed by the recombination of phenotypic traits. The optimization methodology indicated a potential of progression for breeding new cultivars whether aiming for general or specific adaptation. Among these new combinations of phenotypic traits, the best ideotype identified had drought adaptation properties that suited well an "all-purpose" advisory type. In the global context, it could be supposed that the increased phenotypic plasticity of virtual cultivars generated adapted phenotypes in several environments, and thus, that the same genotype is more successful across a range of diverse conditions, as it was illustrated with leaf area index plasticity. When dealing with more local contexts, the increased diversity of virtual cultivars allowed to choose cultivars with distinctive traits that were more adapted to specific conditions. Consequently, cultivar choice reduced the occurrence of worst cases GEM scenarios when working with virtual cultivars whereas it improved only best cases with actual genetic material and this behaviour of the GEM populations could be later analyzed in regard to farmer's risk management.

However, the proposed crop ideotypes are dependent on designs and tools used in computational experiments : the algorithm of the crop model, the description of the target population of environments and that of the virtual cultivars.

As the state of the modeling directly drives the phenotypic prediction, the level of physiological description and the direction of the model development are central. Evans et al. (2013) described different modeling strategies with a typology that associates complexity and aims. In this typology, the SUNFLO model is a *minimal model for a system*, i.e. it uses coarse-grained descriptions that lead to quantitative predictions of essential features of the system. Furthermore, Hammer et al. (2006) indicated that a fine resolution is not always the best starting point. Open-source and open data are also important tools that contribute to improve model intercomparison (Sandve et al., 2013), thus helping choose an adequate representation for the system.

Considering the design of computational experiments, the defined target population of environments was between the level where cultivars are usually advised and sold in France (administrative regions) and the level where cultivars are bred (three large zones in Europe). Targeting a wider environmental variability, e.g. by including some Spanish and Eastern Europe locations, should have an impact on ideotypes diversity. It could be expected that cultivars displaying traits that maximize plant transpiration, i.e. a high potential and duration of leaf area, a weak regulation of water flux (e.g. *ALECH* ideotype), could be ideotypes for wetter Eastern Europe locations.

Secondly, virtual cultivars were designed without taking account of observed correlations between phenotypic traits. This assumption allowed to glimpse the potential gain in productivity permitted by an ideal plant phenotype without taking account of the underlying genetic determinism. Future development of numerical optimization methods could target continuous input values for plant traits and an output function that integrates a feasibility criteria based on phenotypic correlations.

Our approach implies a two step design process for ideotypes : firstly, identify trait combinations that suit the ideotype specification (complex phenotype -> traits), then plan its breeding by resolving the genetic basis of traits combinations (traits -> alleles). However, it is not clear if breeding directly for interesting physiological traits would be less time consuming than integrating stronger genetic determinants in crop models, given that the two methods require a strong effort in phenotyping.

Approaches that link dynamic system modeling (Hammer et al., 2006) with statistical methods used in molecular breeding (Van Eeuwijk et al., 2010) provide a plausible way forward in defining an ideotype as a combination of genetic markers. In such approaches, the development and parameterization of the system model should be integrated with the *-omics* analysis (Poorter et al., 2013; Yin and Struik, 2010). Ideally, high-throughput genotyping and phenotyping (Cobb et al., 2013) should encompass variables from both the real and the modelled system, whether it is to link model parameters to genetic markers (Chenu et al., 2009), to compute biomarkers of state variables from transcriptomic data (Marchand et al., 2013) or to map links between specific genes and morpho-physiological traits (Rengel et al., 2012).

**Dealing with increased complexity at upper scales : toward agroecology**

Genetic progress on sunflower was evaluated at 1.3 % per year, on the 1970-2000 period (Vear et al., 2003), with variable gain thorough years. We quantified the gain provided by educated genotypic choices at 3 to 8 % depending on the genetic material but even if this improvement does not scale with time, we could reasonably suppose that it is added to genetic progress. Furthermore, the exploitation of GEM interactions is often discussed with the perspective of breeding new material (Messina et al., 2011; Vega and Chapman, 2006) but an improved use of actual cultivars guided by ecophysiological knowledge and crop modeling would lead to a modest increase of crop performance. Along with the development of advanced functional breeding methods, this approach is operational on current genetic material and may fit into an ecological transition to a more sustainable agriculture. However, genotypic optimization and breeding seems to act synergistically as improved choices on virtual genetic material allowed both to increase yield expectation and stability, mainly by avoiding worst cases.

The farming system scale changes the optimization problem into a "putting your eggs into many different baskets" one : cultivars displaying different strategies can be allocated to distinct fields, even though the same problem could also be addressed by mixing cultivars in a single plot. In this case, adaptation without the full knowledge of the environmental conditions would allow a better crop performance because of the dampening of the abiotic stress effect at the population level (plants or plots).

The same way that a conciliation of genetics and crop physiology is necessary to bridge the genotype-phenotype gap, the application of computer science to agrosystems is necessary to navigate into this additional layer of complexity created by modeling approaches. Plant biologists therefore suggested to extend systems biology, often focused on the cellular level, to agrosystems (Hammer et al., 2004; Keurentjes et al., 2011; Poorter et al., 2013) to design cropping systems that enhance spatial management of genetic diversity. Analysis frameworks used in such approaches should allow to formulate easily different optimization problems, whether they are discrete (alleles), continuous (traits), constrained (genetic or phenotypic correlations) or multi-objectives ones. Crop systems biology can thus enhance our capability of *in silico* up- and down-scaling between gene expression and crop production (Yin and Struik, 2010).

## Conclusions

In this study, we proposed a methodology relying on dynamic modeling, simulation and optimization to predict phenotypic plasticity and search for desired crop ideotypes. This methodology was applied for the sunflower crop at a country-wide scale to assess whether specific adaptation of cultivars could be considered as a lever to improve crop performance.
Computational experiments show a potential for local adaptation, gradually increasing with knowledge of pedo-climatic conditions. At the regional scale, the impact of adaptation of real cultivars was of the same magnitude as the effect of yearly genetic progress by breeding. With the same analysis applied to virtual cultivars designed by recombining phenotypic traits, adaptation of cultivars had a greater positive impact on crop performance and stability. We argue that such tools could help enhance spatial management of cultivated genetic diversity and evolve into functional breeding approaches.

## Acknowledgements

The authors are grateful to the students (Ewen Gery, Bertrand Haquin, Claire Barbet-Massin) and staff from INRA and ENSAT (Colette Quinquiry, Michel Labarrère, Pierre Maury) and CETIOM (Pascal Fauvin, Philippe Christante, André Estragnat, Emmanuelle Mestries) that helped to constitute the phenotypic database, helped in modeling and simulation (RECORD team from INRA, Helène Raynal, Eric Casellas, Gauthier Quesnel) and provided climatic datasets (AgroClim team from INRA). Grants were provided by the French Ministry of Research (ANR SUNRISE, ANR MICMAC ANR-09-STRA-06) and the French Ministry of Agriculture (AAP CTPS)

# Supplementary data

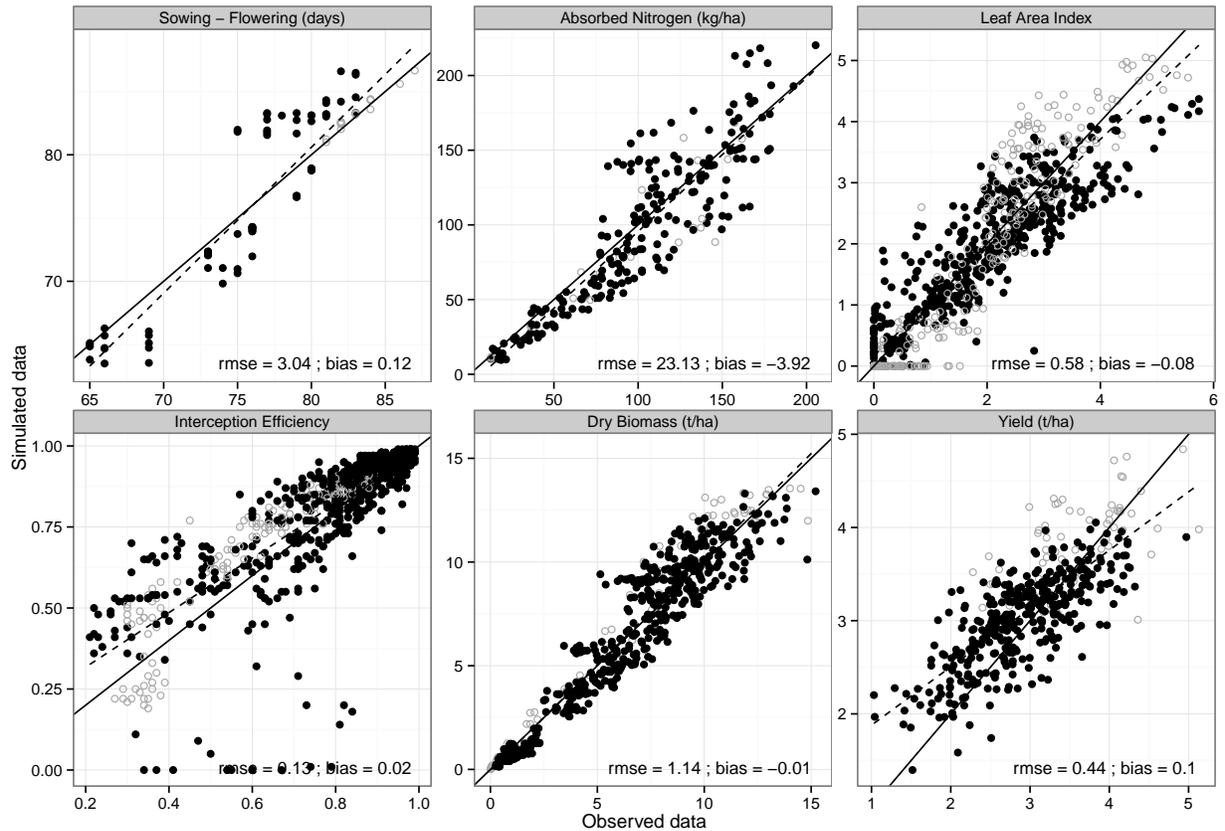

**Supporting figure 1 : Evaluation of model prediction capacity** The prediction capacity of the dynamic model was evaluated on dataset (n=2793) of 6 variables measured in contrasted environmental conditions and different cultivars (see Casadebaig et al. (2011) for trials description). All variables were sampled dynamically along the crop cycle, excepted flowering date and yield. Open symbols represent data used for parameter estimation; solid line represent the 1:1 line; dotted line is a linear regression on all data. The model root mean square error (RMSE) and bias were calculated on independant data points (closed symbols).

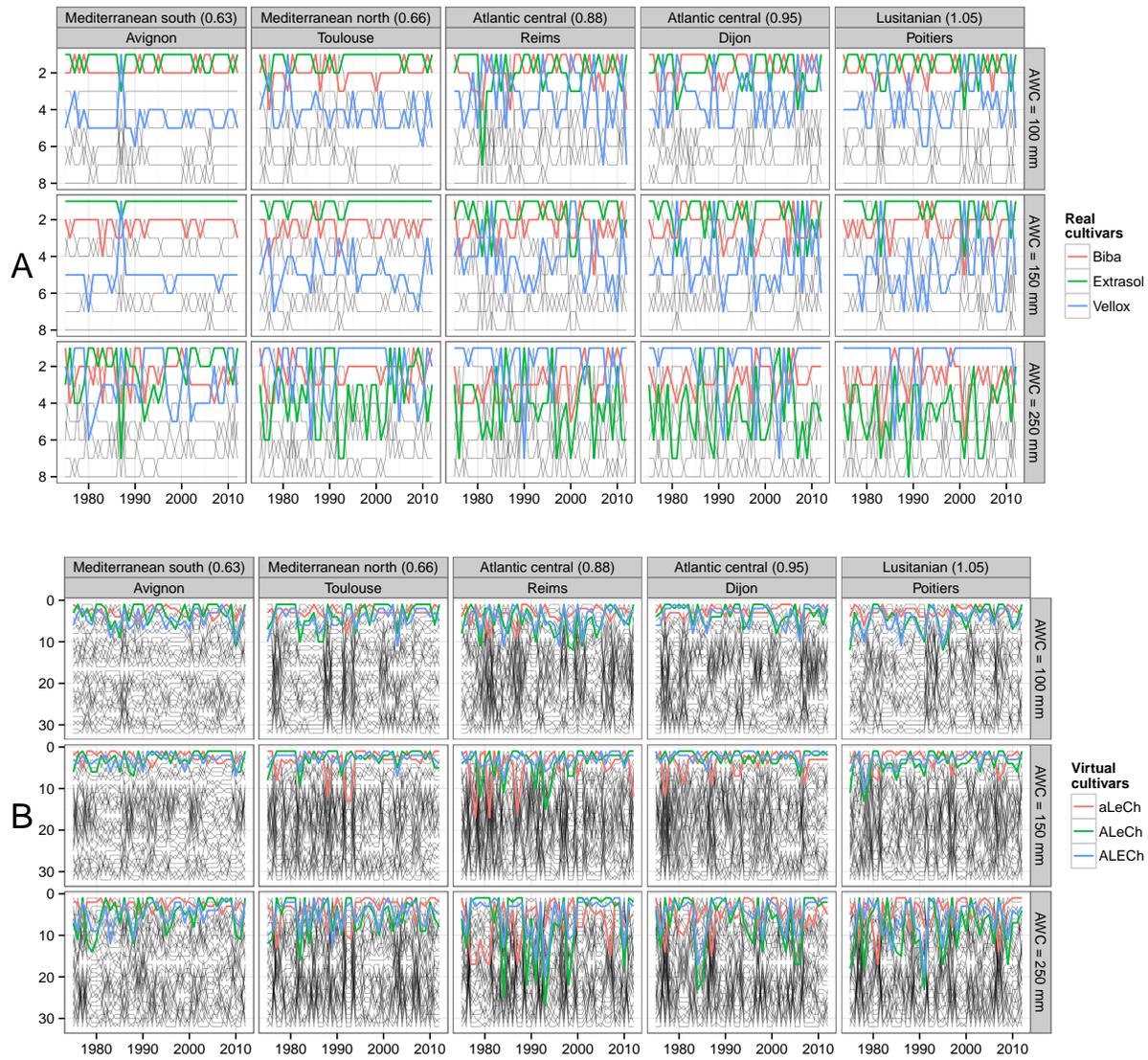

**Supporting figure 2 : Norms of reaction for grain yield illustrate simulated G by E interactions**
Simulated norms of reaction for crop yield are reresented for real cultivars (panel A) and virtual cultivars (panel B). Locations are sorted from left to right according to their aridity index (in brackets) and soils are sorted from top to bottom by increasing available water capacity. In each panel, lines represent the variation of rank of one cultivar according to the years; the overall best cultivars are displayed in color.